\batchmode
\makeatletter
\def\input@path{{/home/runjing_liu/Documents/BNP/genomic_time_series_bnp/writing/nips_workshop_2017/}}
\makeatother
\documentclass[english]{article}\usepackage[]{graphicx}\usepackage[]{color}
\makeatletter
\def\maxwidth{ %
  \ifdim\Gin@nat@width>\linewidth
    \linewidth
  \else
    \Gin@nat@width
  \fi
}
\makeatother

\definecolor{fgcolor}{rgb}{0.345, 0.345, 0.345}

\usepackage{framed}
\makeatletter
 {\par\unskip\endMakeFramed%
 \at@end@of@kframe}
\makeatother

\definecolor{shadecolor}{rgb}{.97, .97, .97}
\definecolor{messagecolor}{rgb}{0, 0, 0}
\definecolor{warningcolor}{rgb}{1, 0, 1}
\definecolor{errorcolor}{rgb}{1, 0, 0}
\newenvironment{knitrout}{}{} 

\usepackage{alltt}
\usepackage[T1]{fontenc}
\usepackage[latin9]{inputenc}
\usepackage{prettyref}
\usepackage{amsmath}
\usepackage{amssymb}
\usepackage[authoryear]{natbib}
\usepackage{xargs}[2008/03/08]

\makeatletter

\providecommand{\tabularnewline}{\\}

\usepackage[final]{nips_2017}
\usepackage{hyperref}
\usepackage{url}

\newcommand{\fig}[1]{Fig.~(\ref{fig:#1})}

\newrefformat{app}{Appendix \ref{#1}}
\newrefformat{eq}{Eq.~(\ref{#1})}
\newrefformat{assu}{Assumption~(\ref{#1})}
\newrefformat{subsec}{Section~(\ref{#1})}
\newrefformat{cor}{Corollary~(\ref{#1})}
\newrefformat{thm}{Theorem~(\ref{#1})}
\newrefformat{def}{Definition~(\ref{#1})}
\newrefformat{prop}{Proposition~(\ref{#1})}
\newrefformat{tab}{Table~(\ref{#1})}

\usepackage{titlesec}

\titleformat{\section}[runin]
{\normalfont\bfseries}
{\S\ \thesection.}{.5em}{}[.\hphantom{2em}]
\titlespacing{\section}
{\parindent}{1.5ex plus .1ex minus .2ex}{0pt}

\makeatother

\usepackage{babel}
\IfFileExists{upquote.sty}{\usepackage{upquote}}{}
\begin{document}

\newcommand{\bnpalpha}{2}
\newcommand{\betamean}{0.38}
\newcommand{\betainfo}{0.10}
\newcommand{\gammascale}{0.10}
\newcommand{\gammashape}{10.00}
\newcommand{\bmean}{0}
\newcommand{\binfo}{0.10}
\newcommand{\nboot}{200}
\newcommand{\kapprox}{30}
\newcommand{\nobs}{1000}
\newcommand{\betadim}{7}
\newcommand{\ntime}{14}
\newcommand{\splinedegree}{3}

\global\long\def\mbe{\mathbb{E}}

\newcommandx\argmin[1][usedefault, addprefix=\global, 1=\eta]{\underset{#1}{\mathrm{argmin}}}

\global\long\def\iid{\stackrel{iid}{\sim}}

\title{Measuring Cluster Stability for Bayesian Nonparametrics Using the Linear Bootstrap}

\author{
Ryan Giordano\thanks{These authors contributed equally.}\\
\texttt{rgiordano@berkeley.edu}
\And
Runjing Liu\textsuperscript{*}\\
\texttt{runjing\_liu@berkeley.edu}
\And
Nelle Varoquaux\textsuperscript{*}\\
\texttt{nelle@berkeley.edu}
\And
Michael I.~Jordan\\
\texttt{jordan@cs.berkeley.edu}
\And
Tamara Broderick\\
\texttt{tbroderick@csail.mit.edu}
}


\maketitle

\section{Introduction}

Clustering is the canonical unsupervised learning problem, in which
we aim to find an assignment of data points to groups, or clusters,
that represent meaningful latent structure in a data set. Bayesian
nonparametric (BNP) models form a particularly popular set of Bayesian
models for clustering due to their flexibility and coherent assessment
of uncertainty. As with any Bayesian model of moderate complexity,
typically the Bayesian posterior cannot be computed exactly for BNP
clustering problems, and an approximation must be employed. Mean-field
variational Bayes (MFVB) forms a posterior approximation by solving
an optimization problem and is widely used due to its speed \citep{blei:2006:dirichletbnp}.
An exact BNP posterior might, at least in theory, vary dramatically
when presented with different data. Certainly we expect small, rare
clusters\textemdash which are ubiquitous in BNP\textemdash to vary
substantially based on the observed data. When reporting the summaries
of the clustering for the purposes of scientific inquiry, it behooves
us to understand how stable, or alternatively how sensitive, this
report is relative to the data \citep{yu:2013:stability}.

If one were to use the bootstrap to assess stability in this analysis
pipeline, it would require a new run of MFVB for each simulated data
set. This time cost is often prohibitively expensive, especially for
exploratory data analyses. We instead propose to provide a fast, automatic
approximation to a full bootstrap analysis based on the infinitesimal
jackknife \citep{jaeckel:1972:infinitesimal,efron:1982:jackknife},
which can be seen as a linear approximation to the global stability
measure provided by the full bootstrap. This locality can buy drastic
time savings, with the infinitesimal jackknife sometimes running orders
of magnitude faster than the bootstrap. We here demonstrate how to
apply this idea to a data analysis pipeline consisting of an MFVB
approximation to a BNP clustering posterior. We show that the necessary
calculations can be done nearly automatically, without tedious derivations
by a practitioner, using modern automatic differentiation software
\citep{maclaurin:2015:autograd}. This automation suggests a generality
to our methods beyond BNP clustering.

In the remainder, we describe the BNP model and MFVB inference in
more detail in \prettyref{sec:Model}. We review summaries for assessing
the output of our clustering, across which we can in turn assess stability,
in \prettyref{sec:cluster_quality}. We describe our new stability
assessment procedure in \prettyref{sec:data_sensitivity}. And we
demonstrate our ability to quickly and accurately quantify stability
in \prettyref{sec:Results} on an application to clustering time-course
gene expression data \citep{shoemaker:2015:ultrasensitive,Luan:2003:clustering}.

\section{Data, Model, and Inference\label{sec:Model}}

Clustering procedures typically estimate which data points are clustered
together, a quantity of primary importance in many analyses. It can
be used to reduce the dimensionality, or to facilitate the interpretation
of complex data sets. For example, genomics experiment often assess
cell activity genome-wide, but many genes behave the same way. Clustering
them thus allows dimensionality reduction that can facilitate interpretation.
Finding robust and stable clusters is thus crucial for appropriate
downstream analysis.

Because the differences and evolution over time of gene expression
yields important insight on gene regulation of the cell-cycle, or
on how cells react to toxins, drugs or viruses, we focus on the specific
task of clustering time course gene-expression data. We use a publicly
available data set of mice gene expression \citep{shoemaker:2015:ultrasensitive},
composed $\ntime$ time points after mice are infected with the influenza
virus. See \prettyref{app:data} for more details.

The observed data consists of expression levels $y_{gt}$ for genes
$g=1,...,n_{g}$ and time points $t=1,...,n_{t}$ (see \fig{basis_graph}
for a single-gene time course data). As described by \citet{Luan:2003:clustering},
we model the time series as a gene-specific constant additive offset
plus a B-spline basis of degree $\splinedegree$ and $\betadim$ degrees
of freedom. We denote the basis matrix by $X$ (see \fig{basis_graph}
in \prettyref{app:data}).

Let $b_{g}$ denote the additive offset for gene $g$, and $y_{g}$
the vector of observations $\left(y_{g1},...,y_{gT}\right)^{\top}$.
Denote the variance of the errors as $\sigma^{2}$ and let $I_{T}$
be the $n_{t}\times n_{t}$ identity matrix. We model each gene's
B-spline coefficients, $\beta_{g}$, using a a stick-breaking representation
of a Bayesian nonparametric (BNP) Dirichlet process mixture model
\citep{ferguson:1973:bayesian,sethuraman:1994:constructivedp}. Excluding
priors, the generative model is then:
\begin{align}
\nu_{k}\vert\alpha\iid Beta\left(1,\alpha\right)\quad\quad & \pi_{k}\left(\nu\right):=\nu_{k}\prod_{j=1}^{k-1}\left(1-\nu_{j}\right)\quad\quad\beta_{k}\iid G_{0}\nonumber \\
z_{g}\vert\nu\iid Mult\left(\pi\left(\nu\right)\right)\quad\quad & \beta_{g}\vert z_{g}=\sum_{k=1}^{\infty}\beta_{k}z_{gk}\quad\quad y_{g}\vert X,\beta_{g},b_{g},\sigma^{2}\iid\mathcal{N}\left(X\beta_{g}+b_{g},I_{T}\sigma^{2}\right)\label{eq:generative_model}
\end{align}
See \prettyref{app:variational} for details of the priors.

For brevity, use the single vector $\theta$ to represent all the
unknown parameters $\nu$, $\beta_{k}$, $z_{gk}$, $\sigma^{2}$,
and $b_{g}$, for all $k$ and $g=1,...,n_{g}$. We are interested
in the posterior $p\left(\theta\vert y\right)$, which is intractable.
To approximate $p\left(\theta\vert y\right)$, we form a variational
approximation to $p\left(\theta\vert y\right)$, denoted $q^{*}\left(\theta\right)$
and parameterized by a real-valued parameter $\eta$, using a truncated
representation of the BNP prior with $K=\kapprox$ components, which
was large enough that more than half of the clusters were essentially
unoccupied \citep{blei:2006:dirichletbnp}. The variational distribution
is chosen as a local minimum of the KL divergence from the true posterior:
\begin{align}
q^{*}\left(\theta\right):=q\left(\theta\vert\eta^{*}\right)\textrm{ where }\eta^{*}:= & \argmin KL\left(q\left(\theta\vert\eta\right)||p\left(\theta\vert Y\right)\right).\label{eq:kl_optimum}
\end{align}
See \prettyref{app:variational} for details of the variational approximation.
Ideally, we would like a global minimum of \prettyref{eq:kl_optimum},
but due to the non-convexity of the problem, we can only guarantee
finding a local minimum. Importantly for the assessment of co-clustering,
knowledge of $\eta^{*}$ allows us to approximate the posterior probability
$\zeta_{gk}\left(\eta^{*}\right):=E_{q^{*}}\left[z_{gk}\right]$,
the posterior probability of gene $g$ belonging to cluster $k$ .
We write $\zeta$ without subscripts to refer to the $n_{g}\times K$
matrix with entries $\zeta_{gk}$.

Finally, we introduce some additional notation related to data sensitivity
that will be useful to describe the bootstrap and the infinitesimal
jackknife in \prettyref{sec:data_sensitivity}. To assess data sensitivity,
we augment our model with scalar per-gene weights, $w_{g}\ge0$, where
$W=\left(w_{1},...,w_{n_{g}}\right)^{\top}$, where we define the
weighted likelihood and corresponding optimal variational parameter:
\begin{align}
\log p\left(Y\vert\theta,W\right) & =\sum_{g=1}^{n_{g}}w_{g}\log p\left(y_{g\cdot}\vert\theta\right)\quad\Rightarrow\quad\eta^{*}\left(W\right):=\argmin KL\left(q\left(\theta;\eta\right)||p\left(\theta\vert Y,W\right)\right).\label{eq:weighted_log_lik_def}
\end{align}
Defining $W_{1}:=\left(1,...,1\right)^{\top}$ we recover the original
variational posterior $\eta^{*}=\eta^{*}\left(W_{1}\right)$. By setting
$W$ to other integer-valued vectors, we can produce the effect of
removing or repeating datapoints, since $p\left(Y\vert\theta\right)$
is exchangeable in $y_{g}$. In particular, by drawing $n_{b}$ bootstrap
weights $W_{b}\sim Multinomial\left(n_{b},n_{b}^{-1}\right)$, for
$b=1,...,n_{b}$, the bootstrap distribution of a function $\phi\left(\zeta\left(\eta^{*}\right)\right)$
can be approximated with the draws $\phi\left(\zeta\left(\eta^{*}\left(W\right)\right)\right)$.
In the remainder of the paper, in a slight abuse of notation, we will
write $\phi\left(W\right)$ in place of $\phi\left(\zeta\left(\eta^{*}\left(W\right)\right)\right)$
below when the meaning is clear from the context.

\section{Clustering stability measures\label{sec:cluster_quality}}

To quantify the stability of a clustering procedure, we must first
define measures of similarity between different clustering outputs.
In particular, we will consider the similarity between the clustering
$\zeta=\zeta\left(W_{1}\right)$, which is clustering at the optimum
$\eta^{*}$, and $\tilde{\zeta}:=\zeta\left(W_{b}\right)$ at bootstrap
weights $W_{b}$. We will use three clustering similarity measures:
the Fowlkes-Mallows index\footnote{Note that we extend the traditional Fowlkes-Mallows index to be a
function of the posterior probabilities $\zeta_{gk}$. See \prettyref{app:clustering_stability_measures}
for more details.} \citep{fowlkes:1983:method}, the normalized mutual information \citep{vinh:2010:clustermutualinformation},
and the bootstrap standard deviation of the elements of the matrix
$\zeta\left(W_{b}\right)$. Let $P\left(k_{1,}k_{2}\right)=\zeta_{k_{1}}^{\top}\zeta_{k_{2}}$,
and denote the entropy of a distribution $H\left(P\right):=\sum_{k}P\left(k\right)\log P\left(k\right)$.
Treating $\zeta$ as fixed, the two similarity measures can be written
as:
\begin{align}
\phi_{FM}\left(\tilde{\zeta}\right)=\frac{\underset{g_{1}g_{2}}{\sum}\left(\zeta_{g_{1}}^{\top}\zeta_{g_{2}}\right)\left(\tilde{\zeta}_{g_{1}}^{\top}\tilde{\zeta}_{g_{2}}\right)}{\sqrt{\underset{g_{1}g_{2}}{\sum}\left(\zeta_{g_{1}}^{\top}\zeta_{g_{2}}\right){}^{2}\cdot\left(\tilde{\zeta}_{g_{1}}^{\top}\tilde{\zeta}_{g_{2}}\right){}^{2}}} & \quad\quad\phi_{MI}\left(\tilde{\zeta}\right)=\frac{\underset{k_{1}k_{2}}{\sum}P\left(k_{1,}k_{2}\right)\log\left(\frac{P(k_{1,}k_{2})}{P(k_{1})\tilde{P}(k_{2})}\right)}{\sqrt{H\left(P\right)H\left(\tilde{P}\right)}}\label{eq:fowlkes_mallows}
\end{align}
See \prettyref{app:clustering_stability_measures} for details. Both
yield scores ranging between 0 and 1, where the higher the scores,
the more similar the clusterings are. Note that while we focus on
these measures, the procedure described below can be applied to any
similarity measures $\phi\left(\tilde{\zeta}\right)$.

\section{Data sensitivity\label{sec:data_sensitivity}}

We now derive a local approximation to the bootstrap using the weight
notation from \prettyref{sec:Model}. Noting that \prettyref{eq:weighted_log_lik_def}
is well-defined even for non-integer values of $W$, and observing
that $KL\left(q\left(\theta;\eta\right)||p\left(\theta\vert Y,W\right)\right)$
is smooth in both $\eta$ and $W$ , it follows that $\eta^{*}\left(W\right)$
is smooth in $W$ in a neighborhood of $W_{1}$. Using the results
from \citet[Appendix D]{giordano:2017:covariances}, and adopting
the shorthand notation $KL\left(\eta,W\right):=KL\left(q\left(\theta;\eta\right)||p\left(\theta\vert Y,W\right)\right)$,
we can then calculate a ``weight sensitivity matrix'' $S$ as
\begin{align}
S:=\left.\frac{d\eta^{*}\left(W\right)}{dW}\right|_{W=W_{1}}= & -\left.\left(\frac{\partial^{2}KL\left(\eta,W\right)}{\partial\eta\partial\eta^{\top}}\right)^{-1}\frac{\partial^{2}KL\left(\eta,W\right)}{\partial\eta\partial W}\right|_{W=W_{1}}.\label{eq:sensitivity_defn}
\end{align}
Although \prettyref{eq:sensitivity_defn} would be tedious to calculate
by hand, it can be calculated exactly using automatic differentiation
in just a few lines of code (see \prettyref{app:optimization} for
more details).

Using $S$, and a single-term Taylor expansion, we can approximate
$\eta^{*}\left(W\right)$ and, in turn, a clustering metric $\phi\left(W\right)$:
\begin{align}
\eta^{*}\left(W\right) & \approx\eta_{Lin}^{*}\left(W\right):=\eta^{*}+S\left(W-W_{1}\right)\label{eq:eta_linear_approximation}\\
\phi\left(W\right)=\phi\left(\zeta\left(\eta^{*}\left(W\right)\right)\right) & \approx\phi_{Lin}\left(W\right):=\phi\left(\zeta\left(\eta_{Lin}^{*}\left(W\right)\right)\right).\label{eq:p_linear_approximation}
\end{align}
Note that the quantities $\zeta$, which are probabilities and must
lie between 0 and 1, can be expected to be extremely non-linear functions
of $\eta^{*}$, but they can be calculated quickly for any given $\eta^{*}$.
We take advantage of this fact to make a linear approximation only
on $\eta^{*}$ rather than calculating $\frac{dp_{gk}}{dW}$ directly.

This is nearly equivalent to the ``infinitesimal jackknife'' of
\citet{jaeckel:1972:infinitesimal} (see also \citet[Chapter 6]{efron:1982:jackknife}),
where $\eta^{*}$ is thought of as a statistic depending on the data
$y_{g}$. The only difference is that we linearize $\eta^{*}$ rather
than the full statistic $\phi$. In order to avoid confusion with
the jackknife estimator of variance, we will refer to $\phi_{Lin}\left(W_{b}\right)$
as the ``linear bootstrap'' in \prettyref{sec:Results} below. Note
that the right-hand side of \prettyref{eq:eta_linear_approximation}
involves only a matrix multiplication once $S$ has been calculated,
but evaluating $\eta^{*}\left(W\right)$ exactly for $W\ne W_{1}$
typically involves re-solving the optimization problem \prettyref{eq:weighted_log_lik_def}.
So although $\phi_{Lin}\left(W_{b}\right)$ is only an approximation,
it can generally be calculated much more quickly than $\phi\left(W\right)$
(as is shown in \prettyref{tab:boot_times} below).

\section{Results\label{sec:Results}}

We optimized \prettyref{eq:weighted_log_lik_def} in Python using
the \texttt{trust-ncg} method of \texttt{scipy-optimize} \citep{scipy}
using an initialization based on K-means. We calculated the necessary
derivatives for the optimization and for \prettyref{eq:sensitivity_defn}
using the automatic differentiation library \texttt{autograd} \citep{maclaurin:2015:autograd}.
See \prettyref{app:optimization} for details.

We first found a high-quality optimum for the original dataset (that
is, at $W=W_{1}$) by choosing lowest KL divergence achieved amongst
200 random restarts\footnote{Our results are not quite as good if we take $\eta^{*}$ to be an
optimum chosen after only 10 rather than 200 initializations \textendash{}
see \prettyref{app:Local-Optima} for more details and discussion. }. We take this optimum to be $\eta^{*}$, the value at which we calculate
the sensitivity $S$ in \prettyref{eq:sensitivity_defn}. Then, for
$n_{b}=\nboot$ different bootstrap weights $W_{b}$, we calculate
three different estimates of $\eta^{*}\left(W_{b}\right)$: ``warm
starts'', $\eta_{Warm}^{*}\left(W_{b}\right)$, which optimize $\eta^{*}\left(W_{b}\right)$
starting at $\eta^{*}$; ``cold starts'', $\eta_{Cold}^{*}\left(W_{b}\right)$
which optimize $\eta^{*}\left(W_{b}\right)$ taking the best of ten
new random K-means initializations, and the linear bootstrap estimates,
which are $\eta_{Lin}^{*}\left(W_{b}\right)$ of \prettyref{eq:eta_linear_approximation}.
For each of these three optima, we compare the bootstrap distribution
of the three stability measures of \prettyref{sec:cluster_quality}.
The median times to calculate each of these measures are given in
\prettyref{tab:boot_times}.

\begin{table}
\begin{centering}
\begin{tabular}{|c||c||c||c|c||c|}
\cline{2-6}
\multicolumn{1}{c||}{} &
$\eta^{*}$ (200 inits) &
$\eta_{Cold}^{*}$ (10 inits) &
$\eta_{Warm}^{*}$ (1 init) &
$\eta_{Lin}^{*}$ (given $S$) &
$S$\tabularnewline
\hline
\hline
Time (s): &
16088 &
931 &
53 &
0.0003 &
12\tabularnewline
\hline
\end{tabular}
\par\end{centering}
\caption{Median times to compute each bootstrap sample (or related quantities)\label{tab:boot_times}}
\end{table}

\fig{cluster_quality} shows the distribution of $\phi_{MI}\left(W_{b}\right)$
and $\phi_{FM}\left(W_{b}\right)$ for the three measures. Although
the bootstrap based on is $\eta_{Lin}^{*}\left(W_{b}\right)$ biased
slightly upwards relative to both of the actual bootstraps, it is
a good approximation to the warm start bootstrap.

\begin{knitrout}
\definecolor{shadecolor}{rgb}{0.969, 0.969, 0.969}\color{fgcolor}\begin{figure}[!h]

{\centering \includegraphics[width=0.98\linewidth,height=0.274\linewidth]{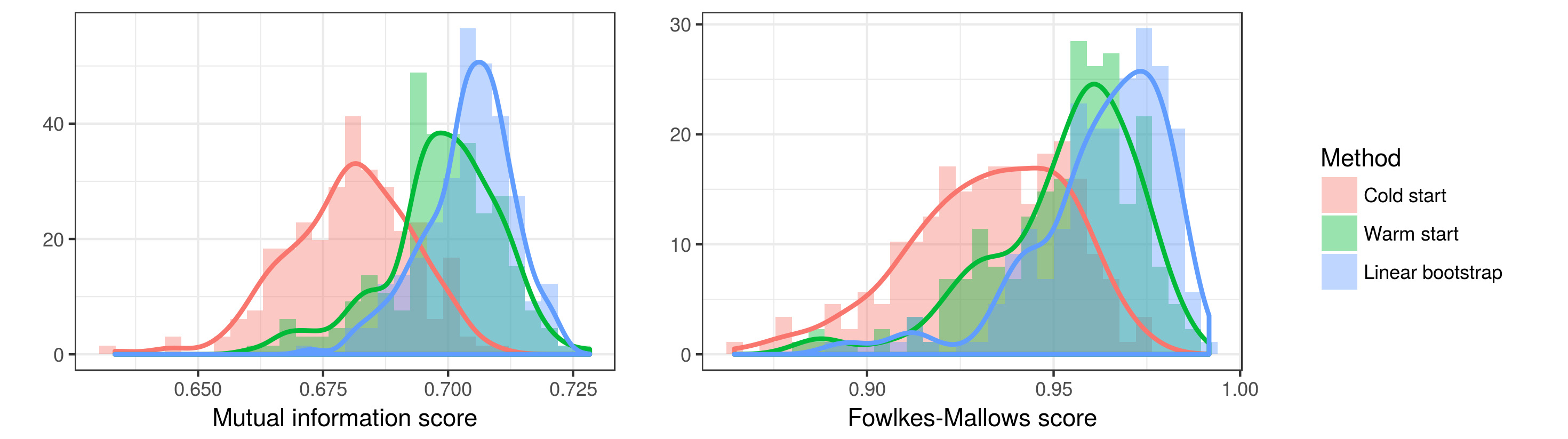}

}

\caption[Cluster quality]{Cluster quality}\label{fig:cluster_quality}
\end{figure}

\end{knitrout}

Finally, we look at the bootstrap standard deviation of the elements
of the matrix $\zeta\left(W_{b}\right)$. \fig{cocluster_sd} shows
the relationship between the co-clustering standard deviation as measured
by $\eta_{Warm}^{*}\left(W_{b}\right)$ on the x-axis and $\eta_{Lin}^{*}\left(W_{b}\right)$
or $\eta_{Cold}^{*}\left(W_{b}\right)$ on the y-axes. Each point
in the graph corresponds to a single value of $W_{b}$, so each graph
contains $B=\nboot$ points. Because the vast majority pairs have
very small standard deviation in both measures of the graph, we condition
on at least one standard deviation being larger than $0.03$. For
both the cold start and the linear bootstrap, most of the mass lies
on the diagonal, indicating a good qualitative correspondence with
the warm start, though there is more frequent extreme deviation in
the linear bootstrap.
\begin{knitrout}
\definecolor{shadecolor}{rgb}{0.969, 0.969, 0.969}\color{fgcolor}\begin{figure}[!h]

{\centering \includegraphics[width=0.98\linewidth,height=0.274\linewidth]{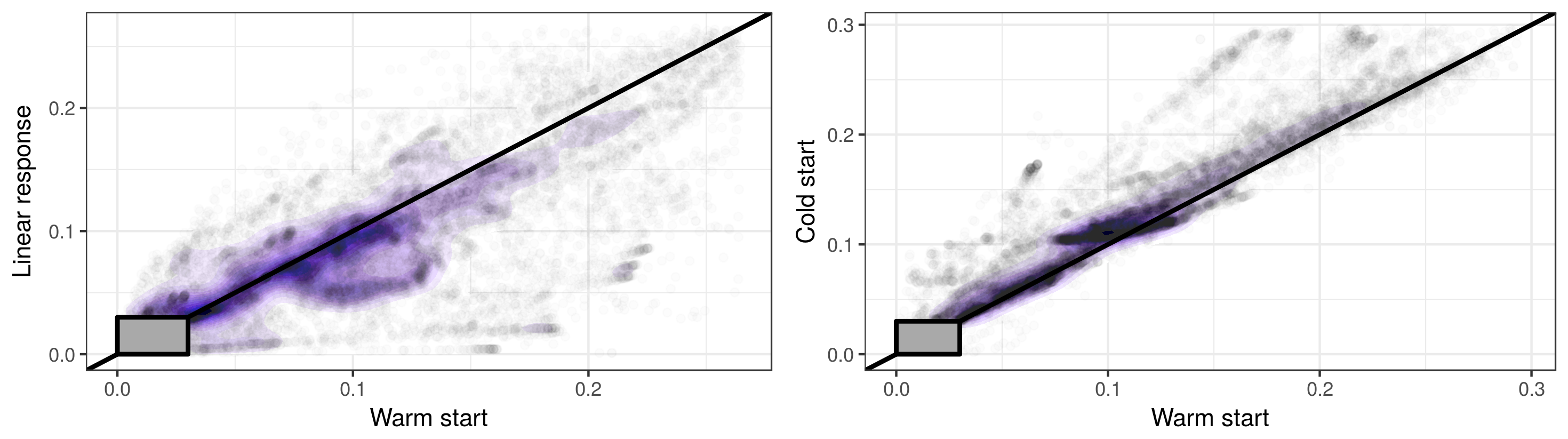}

}

\caption[Standard deviations of elements of the co-clustering matrix for a randomly selected subset of genes]{Standard deviations of elements of the co-clustering matrix for a randomly selected subset of genes. Pairs with standard deviations < 0.03 on both axes are not shown.}\label{fig:cocluster_sd}
\end{figure}

\end{knitrout}

\section{Discussion}

In this work, we studied the stability of time-course gene expression
clustering, using a BNP model and MFVB inference. We compared the
bootstrap, a traditional but computationally intensive approach to
assess stability with a fast, approximate stability assessment procedure,
the linear bootstrap. Instead of re-sampling the data and refitting
the model a large number of times, the linear bootstrap leverages
auto-differentiation tools to obtain a first order approximation of
the re-sampling scheme. We show that the linear bootstrap is a fast
and reasonably accurate alternative to the full bootstrap.

\clearpage{}

\paragraph*{Acknowledgements}

Ryan Giordano and Nelle Varoquaux's research was funded in full by
the Gordon and Betty Moore Foundation through Grant GBMF3834 and by
the Alfred P. Sloan Foundation through Grant 2013-10-27 to the University
of California, Berkeley. Runjing Liu's research was funded by the
NSF Graduate research fellowship. Tamara Broderick's research was
supported in part by a Google Faculty Research Award and the Office
of Naval Research under contract/grant number N00014-17-1-2072.\bibliographystyle{plainnat}
\bibliography{timecourse_clustering}

\pagebreak{}

\appendix

\part*{Appendices}

\section{Data description and processing\label{app:data}}

We use the publicly available mice micro array data set \citep{shoemaker:2015:ultrasensitive}.
Mice were infected with different influenza viruses, and gene expression
was assessed at $\ntime$ time points after infection. We focus on
the influenza virus ``A/California/04/2009'', a mildly pathogenic
virus from the 2009 pandemic season. We normalize the data as described
in \citet{shoemaker:2015:ultrasensitive}. We then apply the differential
analysis tool EDGE between the influenza infected mice and control
mice \citep{Storey:2005:significance}. EDGE yields for each gene
a p-value assessing how differently the genes behave between the two
conditions. We then rank the genes from most significantly differentially
expressed, to least significantly expressed and perform all downstream
analysis on the top $\nobs$ genes.

The observations are unevenly spaced, with more frequent observations
at the beginning. As shown \fig{basis_graph}, each gene also has
multiple measurements at each time point (called biological replicates).
By modeling gene expression as a smooth function, via a B-spline basis,
we naturally model the time aspect of the data, as well as provide
an easy framework for including biological replicates in the clustering.
The reader may observe that the sparse observations at later times
leads to apparent non-smoothness in the fitted time series at late
times, though the B-splines enforce smoothness in actual calendar
time as desired.

\begin{knitrout}
\definecolor{shadecolor}{rgb}{0.969, 0.969, 0.969}\color{fgcolor}\begin{figure}[!h]

{\centering \includegraphics[width=0.98\linewidth,height=0.274\linewidth]{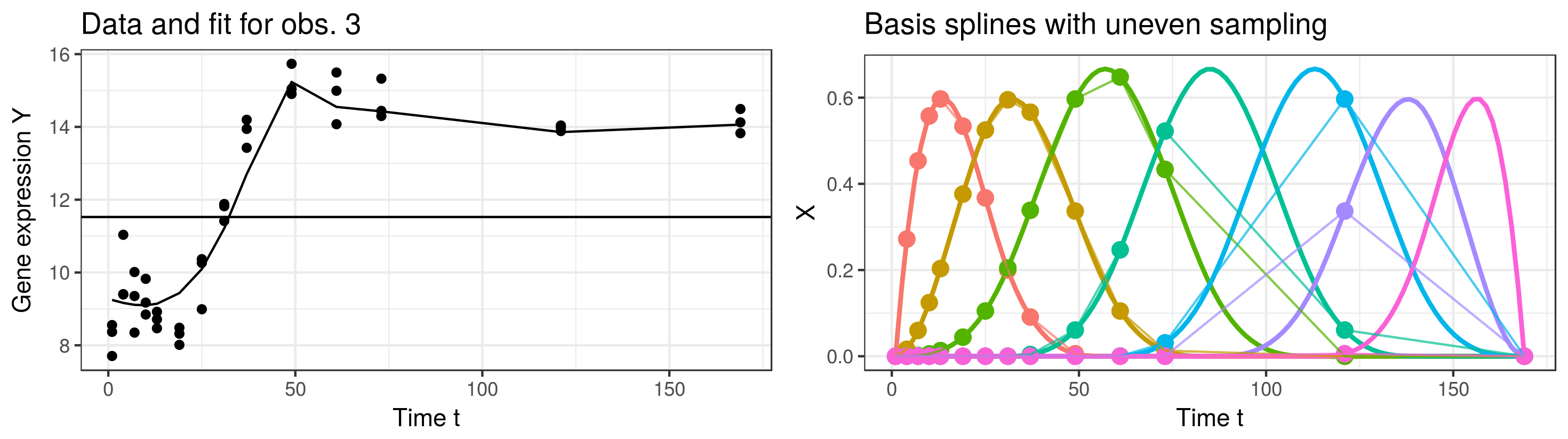}

}

\caption[Data and splines]{Data and splines}\label{fig:basis_graph}
\end{figure}

\end{knitrout}

\section{Variational Inference\label{app:variational}}

We used the following priors:
\begin{align*}
\alpha & =\bnpalpha\\
\beta_{ki} & \iid\mathcal{N}\left(\betamean,\betainfo^{-1}\right)\\
b_{g} & \iid\mathcal{N}\left(\bmean,\binfo^{-1}\right)\\
\tau:=\sigma^{-2} & \sim Gamma\left(\gammascale,\gammashape\right)\textrm{ (shape / scale parameterization)}
\end{align*}

The variational approximation was
\begin{align*}
q\left(\theta\vert\eta\right) & =\delta\left(\beta\right)\delta\left(\tau\right)\prod_{k=1}^{K}\left\{ q\left(\nu_{k}\right)\prod_{g}q\left(z_{gk}\right)q\left(b_{g}\vert z_{gk}=1\right)\right\} ,
\end{align*}
where $\delta\left(\cdot\right)$ denotes a point mass at a parameterized
location \citep{neal:1998:variationalEM} \footnote{Technically, a true point mass does not have a well-defined KL divergence
with respect to the Lebesgue measure on $\beta$ and $\tau$. But
$\delta\left(\beta;\eta_{\beta}\right)$ can be thought of as a density
with constant entropy, and where $\mbe_{\delta}\left[\beta\right]\approx\eta_{\beta}$.
Such a distribution can be approximated arbitrarily closely with a
multivariate normal distribution with vanishing variance, for example.}, $q\left(\nu_{k}\right)$ is a beta distribution, $q\left(z_{gk}\right)$
is a multinomial distribution, $q\left(b_{g}\vert z_{gk}=1\right)$
is a normal distribution, and $\eta$ denotes the vector of parameters
for all these distributions. With this approximation, we seek $\eta^{*}:=\argmin[\eta]KL(q(\theta|\eta)||p(\theta|Y))$.
See Appendix \ref{app:optimization} for details of the optimization.

\section{Optimization \label{app:optimization}}

Note that by parameterizing $q\left(b_{g},z_{g}\right)=\prod_{k=1}^{K}q\left(b_{g}\vert z_{gk}=1\right)q\left(z_{g}\right)$,
the updates for $q\left(b_{g},z_{g}\right)$ have a closed form given
$q\left(\beta,\tau,\nu\right)$. Denote the parameters for $q\left(b_{g},z_{g}\right)$
as $\eta_{local}$ and the parameters for $q\left(\beta,\tau,\nu\right)$
as $\eta_{global}$, and write
\begin{align*}
\hat{\eta}_{local}\left(\eta_{global}\right) & :=\argmin[\eta_{local}]KL\left(\eta_{global},\eta_{local}\right),
\end{align*}
we can write the optimization problem \prettyref{eq:kl_optimum} as
a function of $\eta_{global}$ only:
\begin{align}
\eta_{global} & =\argmin[\eta_{global}]KL\left(\eta_{global},\hat{\eta}_{local}\left(\eta_{global}\right)\right).\label{eq:marginal_kl}
\end{align}
This is valuable because the size of $\eta_{local}$ grows with the
number of genes, but the size of $\eta_{global}$ does not. In addition
to speeding up optimization, \prettyref{eq:marginal_kl} can be easily
differentiated using \texttt{autograd} to calculate the sensitivity
matrix $S$ in \prettyref{eq:sensitivity_defn}.

To solve \prettyref{eq:marginal_kl}, we use a combination of Newton
and quasi-Newton methods. We first choose an initialization by fitting
individual B-splines to each gene expression, and use K-means to cluster
the coefficients; the centroids were used to initialize the variational
means for $\beta_{K}$. From this initialization, we ran BFGS for
300 iterations; at the point where BFGS terminated, we computed the
Hessian of the KL objective, \prettyref{eq:marginal_kl}. This Hessian
was used as a preconditioner for the final Newton trust region steps,
which was iterated to convergence. Hessians were computed using \texttt{autograd}
\citep{maclaurin:2015:autograd}, while BFGS and the newton trust-region
routines were done with the \texttt{BFGS} and \texttt{trust-ncg} methods
of \texttt{scipy-optimize} \citep{scipy}, respectively.

\section{Clustering stability measures \label{app:clustering_stability_measures}}

In this work, we focus on two standard clustering stability measures:
the Fowlkes-Mallows index and the Normalized mutual information. As
mentioned in \ref{sec:cluster_quality}, we adapt the stability measures
to be a function of $\zeta$ and $\tilde{\zeta}$. We here describe
in more details those similarity measures and how we adapted them
for our use case.

First, we will take a closer look at the Fowlkes-Mallows index \citep{fowlkes:1983:method}.
Ignoring for the moment the variational distribution, consider a general
clustering algorithm that outputs binary indicators $z_{gk}$ for
gene $g$ belonging to cluster $k$. Suppose two different runs of
the algorithm (e.g. runs with two different initializations) give
two different outputs $z_{gk}$and $\tilde{z}_{gk}$. Then the Fowlkes-Mallows
similarity index is defined as
\begin{align}
FM=\frac{\sum_{g_{1}g_{2}}C_{g_{1}g_{2}}\tilde{C}_{g_{1}g_{2}}}{\sqrt{(\sum_{g_{1}g_{2}}C_{g_{1}g_{2}}^{2})\cdot(\sum_{g_{1}g_{2}}\tilde{C}_{g_{1}g_{2}}^{2})}}\label{eq:fowlkes_mallows_original}
\end{align}

where $C_{g_{1}g_{2}}:=\sum_{k=1}^{K}z_{g_{1}k}z_{g_{2}k}$ is the
indicator that genes $g_{1}$ and $g_{2}$ are clustered together
under the first clustering; and $\tilde{C}_{g_{1}g_{2}}$ denotes
the same quantity under the second clustering. The numerator in \prettyref{eq:fowlkes_mallows_original}
then counts the number of gene pairs that were co-clustered by both
two clustering results, and the denominator normalizes the index to
be between 0 and 1; hence, values closer to 1 suggest a more similar
clustering.

We modify this definition slightly for our case since we have more
than just binary indicators: we have posterior probabilities for $z_{gk}$
approximated by the variational distribution. This then gives the
probability of co-clustering under the variational distribution, $\mbe_{q^{*}}\left[C_{g_{1}g_{2}}\right]$.
Having two different clustering results now corresponds to having
two different variational distributions for $z$. To measure clustering
similarity here, we simply replace $C_{g_{1}g_{2}}$ and $\tilde{C}_{g_{1}g_{2}}$in
\prettyref{eq:fowlkes_mallows_original} with $\mbe_{q^{*}}\left[C_{g_{1}g_{2}}\right]$
and $\mbe_{\tilde{q}^{*}}\left[C_{g_{1}g_{2}}\right],$ their expectations
under two different variational distributions.

Now, let's turn to the normalized mutual information score. Let $q$
and $\tilde{q}$ be two different variational distributions, with
$\mbe_{q^{*}}\left[z_{gk}\right]:=\zeta_{gk}$ and $E_{\tilde{q}^{*}}\left[z_{gk}\right]:=\tilde{\zeta}_{gkl}$.
Suppose we consider the distribution on labels induced by drawing
a random gene $g$, and then drawing the labels $k_{1}\vert g\sim q\left(z_{g}\right)$
and $k_{2}\vert z_{g}\sim\tilde{q}\left(z_{g}\right)$. Then define
$P\left(k_{1}\right)=\frac{1}{n_{g}}\sum_{g}\zeta_{gk}$, the probability
of cluster $k_{1}$ under the first variational distribution, and
$\tilde{P}\left(k_{2}\right)=\frac{1}{n_{g}}\sum_{g}\tilde{\zeta}_{gk},$
the probability of cluster $k_{2}$ under the second variational distribution;
also let $P(k_{1,}k_{2})=\frac{1}{n_{g}}\sum_{g}\zeta_{gk_{1}}\tilde{\zeta}_{gk_{2}}$,
the joint cluster probabilities. Then the normalized mutual information
score for clustering similarity is given by

\begin{align}
NMI=\frac{\sum_{k_{1}k_{2}}P(k_{1,}k_{2})\log(\frac{P(k_{1,}k_{2})}{P(k_{1})\tilde{P}(k_{2})})}{\sqrt{(\sum_{k}P(k)\log P(k))\cdot(\sum_{k}\tilde{P}(k)\log\tilde{P}(k))}}\label{eq:mut_info_appendix}
\end{align}

The numerator is the mutual information between the two clustering
outputs defined by the variational distributions $q$ and $\tilde{q}$,
with a larger mutual information representing more similar clusterings;
the denominator then normalizes such that the score is between 0 and
1.

\section{Local Optima\label{app:Local-Optima}}

Many unsupervised clustering problems exhibit multiple local optima
in the objective function, even for permutation-invariant quantities
like co-clustering measures, and the problem described in the present
work is no exception. Measures of uncertainty which are based on local
information (like the infinitesimal jackknife) cannot be expected
to capture the frequentist variability due to different initializations
leading to substantively different local optima. The fact that the
cold starts have lower-quality co-clustering than the warm starts
in \fig{cluster_quality} indicates that there exist different local
optima relatively far from $\eta^{*}$. In this section, we briefly
discuss two additional observations concerning local optima.

One might first ask whether the local optima found by the cold start
are much worse than those found by the warm start. The distribution
of KL divergences across the bootstrap samples is shown in \fig{kl_distribution}.
Each point in \fig{kl_distribution} corresponds to two different
estimates at the same weights $W_{b}$, so there are $B=\nboot$ points
in each graph. The linear response KL divergence, which is not evaluated
at an actual optimum, is larger than the corresponding optimal value,
as expected. Note that the cold start KL divergence is not actually
noticeably worse than the warm start KL divergence, suggesting that
there may be meaningful frequentist variability due to local optima
that is not captured by either $\eta_{Warm}^{*}\left(W_{b}\right)$
nor $\eta_{Lin}^{*}\left(W_{b}\right)$.

\begin{knitrout}
\definecolor{shadecolor}{rgb}{0.969, 0.969, 0.969}\color{fgcolor}\begin{figure}[!h]

{\centering \includegraphics[width=0.98\linewidth,height=0.274\linewidth]{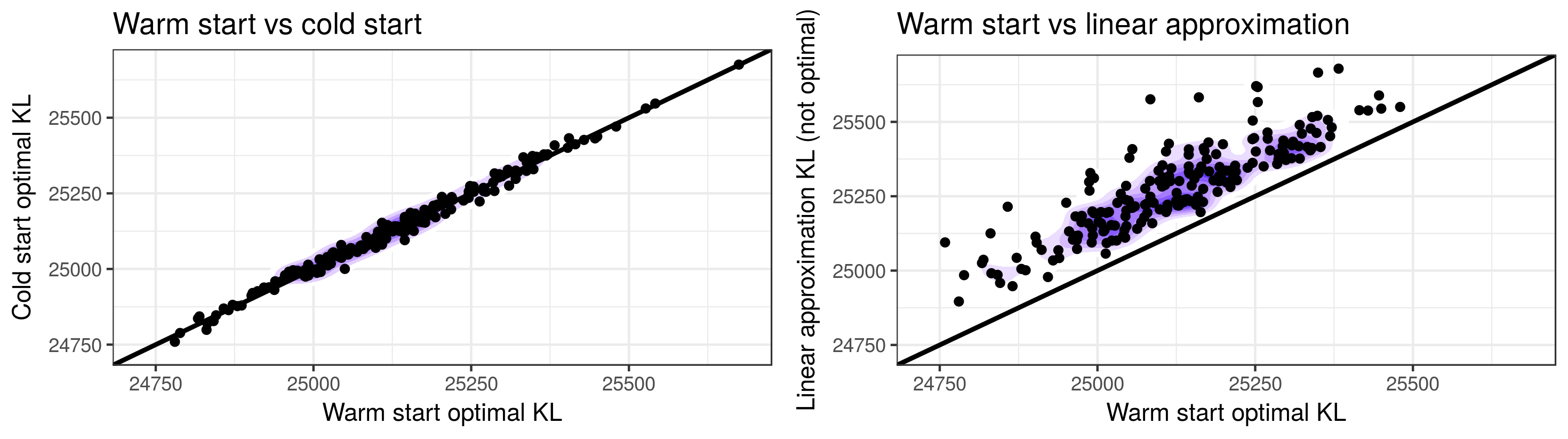}

}

\caption[Distribution of KL divergence relative to the warm start]{Distribution of KL divergence relative to the warm start}\label{fig:kl_distribution}
\end{figure}

\end{knitrout}

Finally, we note that the results in \prettyref{sec:Results} depend
in part on the fact that we are re-starting the optimization in our
bootstrap samples at a high-quality optimum, $\eta^{*}$, chosen as
the best out of 200 random restarts. If, instead, we set $\eta^{*}$
to be the best optimum found after only 10 random restarts, the results
are not quite as good, as seen in \fig{cluster_quality_cold}. This
is probably due both to the base set of cluster assignments, $\zeta$
in \prettyref{eq:fowlkes_mallows}, is not as high-quality an optimum,
and to the fact that optima near $\eta^{*}$, being of lower quality,
is chosen less often during the bootstrap procedure.

\begin{knitrout}
\definecolor{shadecolor}{rgb}{0.969, 0.969, 0.969}\color{fgcolor}\begin{figure}[!h]

{\centering \includegraphics[width=0.98\linewidth,height=0.274\linewidth]{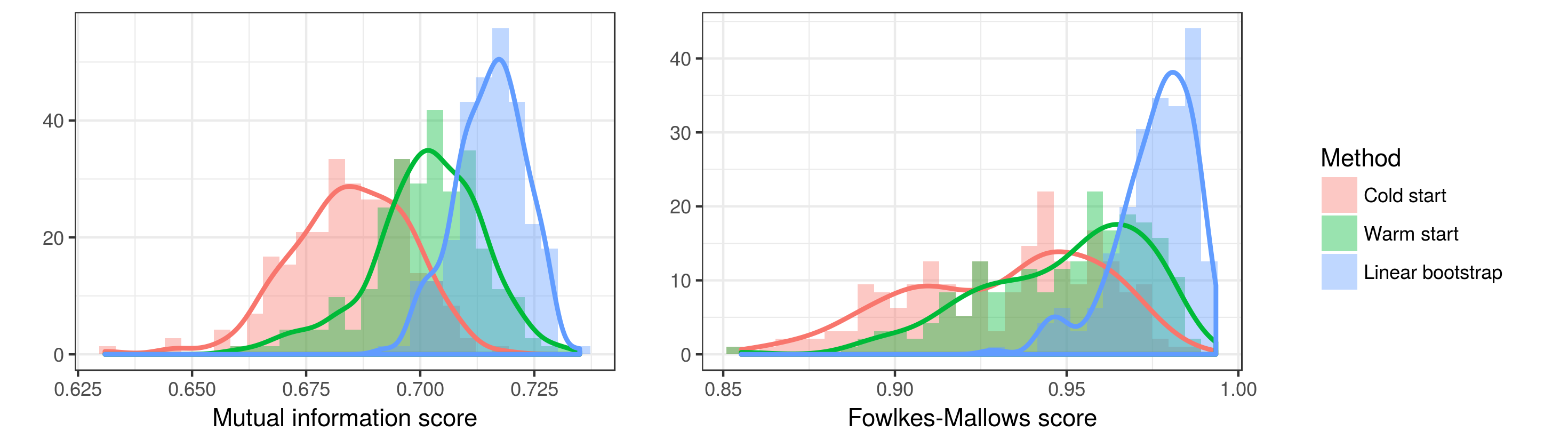}
\includegraphics[width=0.98\linewidth,height=0.274\linewidth]{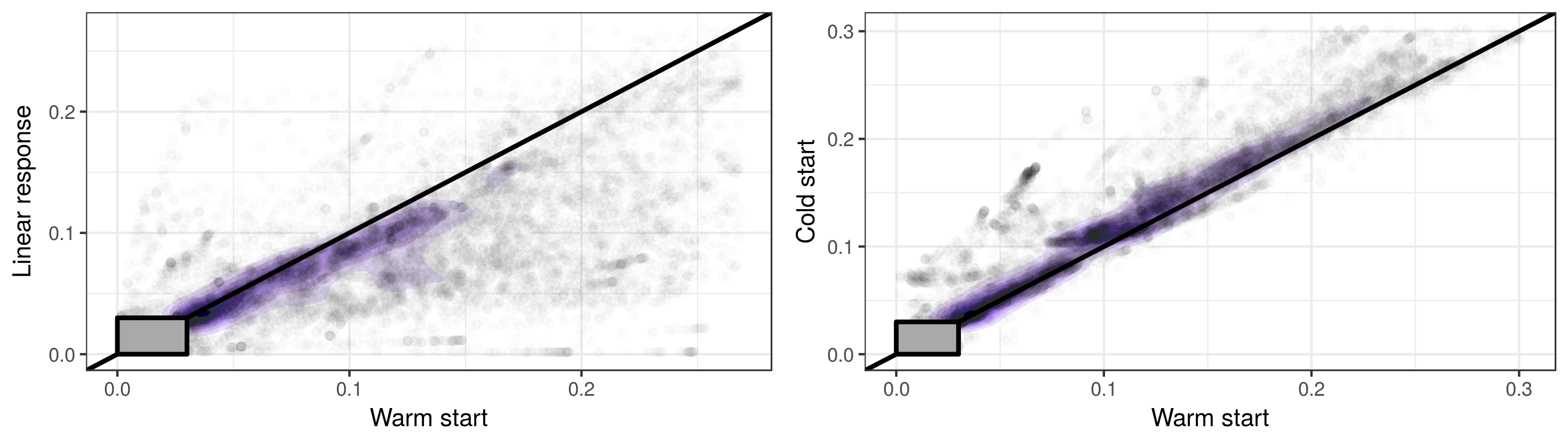}

}

\caption[Results with an initial optimum based on only 10 random restarts rather than 200]{Results with an initial optimum based on only 10 random restarts rather than 200}\label{fig:cluster_quality_cold}
\end{figure}

\end{knitrout}

\end{document}